\documentclass[%
superscriptaddress,
amsmath,amssymb, aps, prc,
floatfix, longbibliography, dvipsnames, twocolumn ]{revtex4-2}

\usepackage{graphicx}
\usepackage{dcolumn}
\usepackage{bm}
\usepackage{orcidlink}
\usepackage{graphics}
\usepackage{amsthm}
\usepackage{amsmath}
\usepackage[ruled,linesnumbered]{algorithm2e}
\theoremstyle{plain}

\usepackage{quantikz}
\usepackage{fontawesome}
\usepackage[caption=false]{subfig}

\usepackage{glossaries}
\newacronym{hep}{HEP}{High Energy Physics}
\newacronym{annni}{ANNNI}{Axial Next-Nearest Neighbor Ising}
\newacronym{qag}{QAG}{Quantum Angle Generator}
\newacronym{qcbm}{QCBM}{Quantum Circuit Born Machine}
\newacronym{qml}{QML}{Quantum Machine Learning}
\newacronym[
	first={Parametrized Quantum Circuit (PQC)},  
	firstplural={Parametrized Quantum Circuits (PQCs)}, 
	plural={PQCs}  
]{pqc}{PQC}{parametrized quantum circuit}  
\newacronym{nisq}{NISQ}{Noisy Intermediate-Scale Quantum}
\newacronym{vqe}{VQE}{Variational Quantum Eigensolver}
\newacronym{mmd}{MMD}{Maximum Mean Discrepancy}
\newacronym{mse}{MSE}{Mean Squared Error}
\glsdisablehyper

\theoremstyle{definition}

\theoremstyle{remark}

\definecolor{maincolor}{RGB}{242,142,0}
\definecolor{secondarycolor}{RGB}{0,166,235}
\definecolor{tertiarycolor}{RGB}{119,119,119}

\begin{document}

\preprint{APS/123-QED}

\title{Classical Pre-Training of Large Shallow Parametrized Quantum Circuits through Pauli Propagation and deployment on Quantum Hardware}
\author{Saverio Monaco \orcidlink{0000-0001-8784-5011}}
\affiliation{RWTH Aachen University, Aachen 52062, Germany}
\affiliation{Deutsches Elektronen-Synchrotron DESY, 22603 Hamburg, Germany}

\author{Jamal Slim \orcidlink{0000-0002-9418-8459}}
\affiliation{Deutsches Elektronen-Synchrotron DESY, 22603 Hamburg, Germany}

\author{Ran Xue \orcidlink{0000-0002-2009-6279}}
\affiliation{Deutsches Elektronen-Synchrotron DESY, 22603 Hamburg, Germany}

\author{Dirk Kr\"ucker \orcidlink{0000-0003-1610-8844}}
\affiliation{Deutsches Elektronen-Synchrotron DESY, 22603 Hamburg, Germany}

\author{Kerstin Borras \orcidlink{0000-0003-1111-249X}}
\affiliation{RWTH Aachen University, Aachen 52062, Germany}
\affiliation{Deutsches Elektronen-Synchrotron DESY, 22603 Hamburg, Germany}

\date{\today}

\begin{abstract}
	Scaling parametrized quantum circuits to a larger number of qubits is a
	central challenge in Quantum Machine Learning: gradient estimation via the
	parameter-shift rule quickly becomes prohibitively expensive, while
	gradient-free optimizers offer limited convergence guarantees. We apply the
	Pauli propagation framework to a broad class of Quantum Machine Learning
	problems in which the loss is expressed as a function of expectation values
	of the circuit's observables, while keeping the relevant quantity of
	interest, such as the quantum state or its samples, classically intractable
	to compute. By propagating the observables backward through the circuit, we
	obtain an explicit parametric expression for the loss that can be minimized
	entirely on classical hardware. We also present complementary strategies to
	make propagation tractable at large scale, including weight-based and
	degree-based truncations and pruning heuristics. We demonstrate the
	approach on two benchmarks run on 64-qubit circuits: Variational Quantum
	Eigensolver on a two-dimensional Ising model, and training of a
	discrete-type generative model for binary images. In both cases, the
	classically pre-trained parameters are deployed on the IBM quantum processor
	\textit{ibm\_berlin}. For generative learning, we additionally show that the
	circuits considered here require substantially fewer parametrized gates than
	IQP-style ansätze, reducing transpilation overhead and making deployment on
	current quantum hardware feasible.
\end{abstract}

\maketitle

\section{Introduction}
\label{sec:intro}

\Gls{qml} algorithms have in recent years been under careful investigation,
asking whether \glspl{pqc} (either standalone or within hybrid classical-quantum
architectures) may yield an advantage over classical neural networks in
generalization~\cite{Monaco_2023,caro2022generalization}, parameter efficiency,
or training convergence.
Meaningfully testing such an advantage, however,
requires operating at a scale beyond classical simulability (roughly $\ge$30
qubits) as below that threshold, a classical simulator can exactly reproduce the
quantum model's behavior.

Quantum hardware capable of operating at such scale is already available, with
current IBM quantum computers exceeding 100 qubits~\cite{ibmroadmap}. However,
hardware availability does not straightforwardly enable extensive training.
Gradient estimation via the parameter-shift rule requires two circuit
evaluations per parameter~\cite{paramshift,schuld2019evaluating}, becoming prohibitively expensive at
large scale. Gradient-free optimizers such as COBYLA~\cite{cobyla} or
SPSA~\cite{spsa} reduce the number of evaluations but offer weaker convergence
guarantees and still demand repeated hardware access. A more structured approach
introduces families of parametrized circuits for which gradient estimation
requires significantly fewer circuit evaluations~\cite{backpropscaling};
however, this comes at the cost of restricting the ansatz architecture. The most
compelling approach to date was proposed in the paper: ``\textit{Train on
	classical, deploy on quantum: scaling generative quantum machine learning to a
	thousand qubits}''~\cite{trainonclassical}, where the authors fully offload
training to classical resources.

Building on the same philosophy, we extend classical training to a
broader class of \glspl{pqc}, namely \textit{locally scrambling}
circuits~\cite{classicallyestimating}, for \gls{qml} problems in which the loss
is expressed as a function of expectation values of observables. Using Pauli
propagation, we derive the explicit functional form of these expectation values
as a function of the circuit parameters $\langle O \rangle(\boldsymbol{\theta})
	\approx f(\boldsymbol{\theta})$. While such surrogates have previously been
employed to scan expectation
landscapes~\cite{fontana2023classical,rudolph2023lowesa,lerch2024patches}, here
$f(\boldsymbol{\theta})$ serves as a training objective minimized through
standard classical optimization, fully offloading training from the quantum
device. We
demonstrate this approach on systems of 64 qubits, well beyond the statevector
simulability regime, and validate the resulting models by deploying
them on actual quantum hardware.

The paper is organized as follows. In Sec.~\ref{sec:prop} we introduce the Pauli
propagation framework, deriving the generic form of a propagated observable and
discussing the limits of exact propagation. We then present two complementary
strategies to make propagation tractable at large scale: weight- and
degree-based truncation schemes and pruning strategies. In
Sec.~\ref{sec:apps} we demonstrate the approach on two \gls{qml} benchmarks: a
\gls{vqe} task and a quantum generative model for binary image generation, both
conducted on 64-qubit circuits and validated on the IBM quantum processor
\textit{ibm\_berlin}. Finally, Sec.~\ref{sec:conc} offers conclusions and an
outlook on future directions.

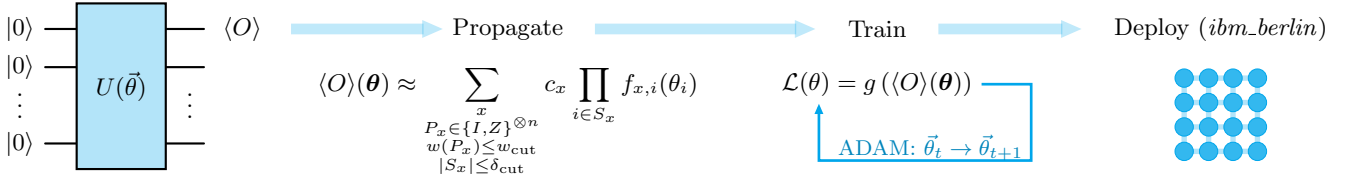
\begin{figure*}[!t]
	\resizebox{\linewidth}{!}{\begin{tikzpicture}[
	connection/.style={
	->,
	line width=1pt,
	secondarycolor,
	>={Triangle[length=4,width=3]}
	},
	fadedconnection/.style={
	->,
	line width=4pt,
	secondarycolor,
	>={Triangle[length=8,width=6]},
	opacity=.25
	},
	]

	\node (ansatzO) at (0,0) {
		\begin{quantikz}[row sep={0.5cm,between origins}, wire types={n,n,n,n}]
			\lstick{$|0\rangle$}                  & [-.25cm] \gate[4, style={fill=secondarycolor!30}]{\,\, U(\vec{\theta}) \,\,}\qw & \qw                              & [-.5cm] \rstick{\,\,$\langle O\rangle$} \\
			\lstick{$|0\rangle$}                  & \qw                                                                             & \qw                              &                                         \\
			\lstick{\raisebox{.2cm}{\vdots}}\quad &                                                                                 & \lstick{\raisebox{.2cm}{\vdots}} &                                         \\
			\lstick{$|0\rangle$}                  & \qw                                                                             & \qw                              &
		\end{quantikz}
	};

	\node (prop) at ([xshift=3cm, yshift=.75cm]ansatzO.east) {
		Propagate
	};
	\node (propformula) at ([yshift=-.95cm]prop.south) {
	$\langle O \rangle(\boldsymbol{\theta})
		\approx \displaystyle\sum_{\substack{x\\ P_x \in \{I,Z\}^{\otimes n} \\ w(P_x) \leq w_{\mathrm{cut}}\\ |S_x| \leq \delta_{\mathrm{cut}}}}
		c_x \prod_{i \in S_x} f_{x,i}(\theta_i)$
	};

	\node (loss) at ([xshift=4cm]prop.east) {
		Train
	};
	\node (lossfun) at ([yshift=-.5cm]loss.south) {
		$\mathcal{L}(\theta) = g\left(\langle O \rangle(\boldsymbol{\theta})\right)$
	};
	\draw[connection] (lossfun.east) -- ([xshift=.65cm]lossfun.east) -- ([xshift=.65cm, yshift=-1cm]lossfun.east) -- ([xshift=.6cm, yshift=-1cm]lossfun.west) -- ([xshift=.6cm, yshift=-.25cm]lossfun.west);
	\node (losslabel) at ([xshift=.7cm, yshift=-1.33cm]loss.south) {
		\color{secondarycolor!80!black} \footnotesize ADAM: $\vec\theta_t \to \vec\theta_{t+1}$
	};

	\node (deploy) at ([xshift=4cm]loss.east) {
		Deploy (\textit{ibm\_berlin})
	};
	\node[scale=.4] (ibmberlin) at ([yshift=-.85cm]deploy.south) {
		\begin{tikzpicture}[
				every node/.style={
						circle, draw, inner sep=0pt,
						minimum size=0.6cm,
						font=\sffamily\small\bfseries, text=secondarycolor!60,
					},
				lightnode/.style={fill=secondarycolor!20, draw=secondarycolor!50, line width=0.8pt},
				node/.style={fill=secondarycolor!80, draw=secondarycolor, line width=0.8pt, text=white},
				edge/.style={draw=secondarycolor!30, line width=5pt},
			]

			\def\dx{0.8}
			\def\dy{-0.8}
			\foreach \row in {0,...,3}{
					\foreach \col in {0,...,2}{
							\pgfmathtruncatemacro{\a}{\row*10 + \col}
							\pgfmathtruncatemacro{\b}{\row*10 + \col + 1}
							\draw[edge] (\col*\dx, \row*\dy) -- (\col*\dx+\dx, \row*\dy);
						}
				}
			\foreach \row in {0,...,2}{
					\foreach \col in {0,...,3}{
							\pgfmathtruncatemacro{\a}{\row*10 + \col}
							\pgfmathtruncatemacro{\b}{\row*10 + \col + 10}
							\draw[edge] (\col*\dx, \row*\dy) -- (\col*\dx, \row*\dy+\dy);
						}
				}


			\foreach \row in {0,...,3}{
					\foreach \col in {0,...,3}{
							\pgfmathtruncatemacro{\n}{\row*10 + \col}
							\pgfmathsetmacro{\x}{\col*\dx}
							\pgfmathsetmacro{\y}{\row*\dy}
							\node[node] at (\x,\y) { };
						}
				}
		\end{tikzpicture}
	};

	\draw[fadedconnection] ([xshift=-2cm]prop.west) -- (prop.west);
	\draw[fadedconnection] ([xshift=.3cm]prop.east) -- ([xshift=-.3cm]loss.west);
	\draw[fadedconnection] ([xshift=.3cm]loss.east) -- ([xshift=-.3cm]deploy.west);
\end{tikzpicture}} \caption{Outline of the
		proposed algorithms. An ansatz and an observable are first defined. The
		observable is then propagated through the circuit, yielding a function
		of the circuit's parameters, returning an approximation of the
		expectation value of the observable. A loss is constructed from this
		function and minimized classically via Adam, producing a set of
		optimized parameters. The ansatz with the trained parameters can then be
		deployed on actual quantum hardware.}
	\label{fig:outline}
\end{figure*}

\section{Propagation of Observables}
\label{sec:prop}

In quantum computation, the \emph{Schr\"{o}dinger
picture} is the natural description of circuit
execution: an initial state $|\mathbf{0}\rangle$ is evolved forward through a
sequence of unitaries $U = U_L \cdots U_1$, yielding the final state
\mbox{$|\psi(\boldsymbol{\theta})\rangle =
		U(\boldsymbol{\theta})|\mathbf{0}\rangle$}, whose measurement statistics can be
used to estimate expectation values. An equivalent but computationally distinct
viewpoint is the \emph{Heisenberg picture}~\cite{heisenberg}, in which the state
is held fixed at $|\mathbf{0}\rangle$ and the observable $O$ is evolved backward
through the circuit via the similarity transformation
\begin{equation}
	O'(\boldsymbol{\theta}) \coloneqq U^\dagger(\boldsymbol{\theta})\, O\, U(\boldsymbol{\theta}).
\end{equation}
This backward evolution is the starting point of \emph{Pauli
	propagation}~\cite{angrisani2025simulating, heisenberg}. Because any
observable can be decomposed in the $n$-qubit Pauli basis $\mathcal{P}_n =
	\{I, X, Y, Z\}^{\otimes n}$, it suffices to track how each Pauli string $P
	\in \mathcal{P}_n$ transforms under conjugation by every gate. Two
quantities characterize a Pauli string as it propagates. The \emph{weight}
$w(P)$ is the number of non-identity single-qubit factors, i.e.\ the number
of qubits on which $P$ acts non-trivially; a weight-1 string acts on a
single qubit while a weight-$n$ string acts on all $n$ qubits. The
\emph{trigonometric degree} of a term in the propagated observable is the number of
trigonometric factors ($\sin$ or $\cos$) it has accumulated, equal to the
number of parametrized rotation gates that contributed a branching step
along its Pauli
path~\cite{monaco2025symbolicpaulipropagationgradientenabled}. Following are
the main families of gates and how they affect these two quantities.

\paragraph{Single-qubit Clifford gates.}
Single-qubit Clifford gates (e.g.\ $H$, $S$) permute the single-qubit Paulis,
mapping one string to exactly one other string of the same weight. For example,
the Hadamard gate swaps the $X$ and $Z$ bases: $H^\dagger X H = Z$ and
$H^\dagger Z H = X$, while $Y$ maps to $-Y$. No branching or weight change
occurs.

\paragraph{Two-qubit Clifford gates.}
Two-qubit Clifford gates such as $\mathit{CNOT}$ and $CZ$ also map Pauli strings
to single Pauli strings without branching, but they can \emph{increase the
	weight} by spreading support across both qubits. For example,
$\mathit{CNOT}^\dagger\, (X \otimes I)\, \mathit{CNOT} = X \otimes X$: a
weight-1 Pauli word becomes weight-2. The weight can grow by at most one per
two-qubit gate, so the total weight is bounded by the number of entangling
gates.

\paragraph{Non-parametrized non-Clifford gates.}
A non-Clifford gate with a fixed angle (e.g.\ the $T$ gate) maps a Pauli string
to a pair of Pauli strings with fixed real coefficients. For example, $T^\dagger
	X T = \tfrac{1}{\sqrt{2}} X - \tfrac{1}{\sqrt{2}} Y$. The coefficients are
constants determined by the gate angle, so no free parameter is introduced but
the number of terms in the sum doubles for each anticommuting gate encountered.

\paragraph{Parametrized rotations.}
A single-qubit Pauli rotation $R_P(\theta) = e^{-i\theta P/2}$ conjugates a
Pauli string $Q$ as
\begin{equation}
	\label{eq:rotprop}
	R_P^\dagger(\theta)\, Q\, R_P(\theta)
	=
	\begin{cases}
		Q                                     & \text{if } [P, Q] = 0,   \\
		\cos(\theta)\, Q + i\sin(\theta)\, PQ & \text{if } \{P, Q\} = 0.
	\end{cases}
\end{equation}
When $P$ and $Q$ anticommute, the gate \emph{branches} the Pauli string into two
terms: a cosine branch retaining $Q$ and a sine branch producing $PQ$. For a
single-qubit rotation, $PQ$ has the same weight as $Q$, so the weight changes
only at multi-qubit gates. Crucially, the coefficients now depend on the free parameter
$\theta$, so the full propagated observable accumulates trigonometric factors at
every branching gate, yielding a sum of products of trigonometric terms. The
number of branches grows exponentially with the number of anticommuting
parametrized gates encountered during propagation, making exact propagation
intractable in general and motivating the strategies explained following in
Sec.~\ref{subsec:truncs} and Sec.~\ref{subsec:pruns}.

Collecting all branches produced by the full circuit, the propagated observable
takes the form
\begin{equation}
	\label{eq:propagated_obs}
	O'(\boldsymbol{\theta}) = \sum_{x} c_x \prod_{k \in S_x} f_{x,k}(\theta_k)\; P_x,
\end{equation}
where the sum runs over all \emph{Pauli paths} $x$, $P_x \in \mathcal{P}_n$ is
the Pauli string reached at the end of path $x$, $c_x \in \mathbb{R}$ is a real
coefficient, $S_x \subseteq \{0,\ldots,p-1\}$ is the set of parametrized gates
that branched along path $x$, and $f_{x,k} \in \{\sin, \cos\}$ is the
trigonometric factor contributed by gate $k$. The trigonometric degree of path $x$ is
$|S_x|$, i.e.\ the number of trigonometric factors it carries.

The expectation value of the propagated observable in the initial state
$|\mathbf{0}\rangle = |0\rangle^{\otimes n}$ then gives the loss:
\begin{equation}
	\langle O \rangle(\boldsymbol{\theta})
	= \langle \mathbf{0} | O'(\boldsymbol{\theta}) | \mathbf{0} \rangle
	= \sum_{x} c_x \prod_{k \in S_x} f_{x,k}(\theta_k)\;
	\langle \mathbf{0} | P_x | \mathbf{0} \rangle.
\end{equation}
The key simplification is that $\langle \mathbf{0} | P_x | \mathbf{0} \rangle$
is non-zero only when $P_x$ is composed exclusively of $Z$ and $I$ operators:
$\langle 0|I|0\rangle = \langle 0|Z|0\rangle = 1$, while $\langle 0|X|0\rangle =
	\langle 0|Y|0\rangle = 0$. Any path $x$ whose terminal Pauli string $P_x$
contains at least one $X$ or $Y$ factor contributes nothing to the expectation
value and can be discarded. This provides a natural filter that eliminates a
large fraction of paths without any approximation, and reduces the full sum to
\begin{equation}
	\langle O \rangle(\boldsymbol{\theta})
	= \sum_{\substack{x \\ P_x \in \{I,Z\}^{\otimes n}}}
	c_x \prod_{k \in S_x} f_{x,k}(\theta_k).
\end{equation}

For a step-by-step example showing how a Pauli string branches into the full
propagated expression, see the appendix of
\cite{monaco2025symbolicpaulipropagationgradientenabled}.

\subsection{Truncations}
\label{subsec:truncs}

The sum in Eq.~\eqref{eq:propagated_obs} runs over exponentially many Pauli
paths in general, making exact propagation
intractable~\cite{angrisani2025simulating,
	monaco2025symbolicpaulipropagationgradientenabled}. Both the weight $w(P_x)$ and
the trigonometric degree $|S_x|$ of a path provide natural measures of its computational
cost and, as argued below, of its likely contribution to the expectation value.
Setting cutoffs on these two quantities yields two complementary approximation
schemes that together make propagation tractable at scale.

\paragraph{Weight truncation.}
Each two-qubit gate can increase the weight of the active Pauli string by one,
so paths that pass through many such gates tend to reach higher weights. It has
been shown~\cite{classicallyestimating} that, for locally scrambling
ans\"{a}tze and on average over the circuit parameters, the contribution of a
path to the expectation value decreases exponentially with its terminal weight
$w(P_x)$. The applicability of this result to the ansatz used here is discussed
in Sec.~\ref{sec:applications}. We
exploit this by fixing a cutoff $w_{\mathrm{cut}}$ and discarding any path whose
active Pauli string exceeds this weight during propagation, preventing it from
branching further. Because heavier strings are truncated before they produce
descendants, this limits the total number of terms that must be tracked and
constitutes the primary control on computational cost.

\paragraph{Trigonometric-degree truncation.}
Each branching step from a parametrized gate increases the trigonometric degree of a path
by one: the anticommuting arm picks up a factor of $\sin(\theta_k)$ while the
commuting arm picks up $\cos(\theta_k)$, so a path of trigonometric degree $|S_x|$ carries
a coefficient that is a product of $|S_x|$ sines and cosines. Since
$|\sin(\theta)|\leq 1$ and $|\cos(\theta)|\leq 1$, a high-degree term is the
product of many such factors, its average contribution to the expectation value
decays exponentially with
trigonometric degree~\cite{monaco2025symbolicpaulipropagationgradientenabled} (under
uniformly-distributed parameters assumption). Setting a trigonometric-degree cutoff
$\delta_{\mathrm{cut}}$ and discarding all paths with $|S_x|>\delta_{\mathrm{cut}}$
provides a second independent control on the complexity of the propagated
observable, complementing the weight truncation above. Applying both cutoffs
simultaneously, the truncated approximation of the loss becomes
\begin{equation}
	\label{eq:truncated_obs}
	\langle O \rangle(\boldsymbol{\theta})
	\approx \sum_{\substack{x \\ P_x \in \{I,Z\}^{\otimes n} \\ w(P_x) \leq w_{\mathrm{cut}} \\ |S_x| \leq \delta_{\mathrm{cut}}}}
	c_x \prod_{k \in S_x} f_{x,k}(\theta_k).
\end{equation}

\subsection{Prunings}
\label{subsec:pruns}
Truncations introduced above are approximations: they discard terms that are
small but generally non-zero. Pruning is conceptually distinct: a pruned branch
is one that is \emph{provably} guaranteed to contribute zero to the final
expectation value $\langle
	\mathbf{0}|O'(\boldsymbol{\theta})|\mathbf{0}\rangle$, regardless of the
parameters. Removing such branches is therefore exact, introducing no error
whatsoever.

Pauli propagation can be naturally viewed as a binary tree search problem. Each
parametrized gate that anticommutes with the active Pauli string forks the
current branch into two children (Eq.~\eqref{eq:rotprop}). The full tree of
paths corresponds to the sum in Eq.~\eqref{eq:propagated_obs}. A pruner
identifies, at an early node of the tree, that no leaf reachable from that node
can survive the zero-bracket filter $\langle\mathbf{0}|P_x|\mathbf{0}\rangle
	\neq 0$, and discards the entire subtree before it is expanded. Since discarded
subtrees may themselves be exponentially large, early pruning can yield dramatic
reductions in propagation cost.

\paragraph{Dead-qubit pruner.}
A Pauli word contributes to the expectation value only if every qubit carries
$Z$ or $I$ at the end of propagation. A gate can change the operator on a qubit
only if that qubit appears in the gate's wire list. Therefore, if a qubit
currently carries $X$ or $Y$ and no remaining gate touches that qubit, its
operator is permanently frozen in a non-$ZI$ state: the word can never reach a
non-zero bracket and is safely discarded. Concretely, the pruner precomputes,
for each propagation step $t$, the set $\mathcal{A}_t$ of qubits touched by any
gate from step $t$ onward. A word is pruned at step $t$ if it carries $X$ or $Y$
on any qubit outside $\mathcal{A}_t$.

\paragraph{XY-weight pruner.}
For every gate in our library (single-qubit gates, $\mathit{CNOT}$, $CY$, $CZ$,
$\mathit{SWAP}$ and controlled rotations), a gate can reduce the XY-weight of a
Pauli word (the number of qubits carrying $X$ or $Y$) by at most one; for a
general $k$-qubit gate this budget becomes $k$. A word whose XY-weight exceeds the number
of remaining gates that intersect its causal cone can therefore never reach
XY-weight zero, and is discarded.

Both pruners are activated at every propagation step and operate in-place on the
active observable, adding negligible overhead while potentially eliminating
large portions of the tree before they are explored.

The two strategies described above are general-purpose within the supported
gate set: they rely only on the zero-bracket condition for any circuit topology. In practice, however, they tend
to trigger most strongly near the end of the circuit, where few gates remain and
the causal cones are small. Pruning earlier in the propagation, where it would
be most valuable, as it cuts off exponentially larger subtrees, requires
exploiting additional structure: depending on symmetries of the circuit
architecture or of the observable~\cite{teng2026leveragingsymmetrymergingpauli},
it may be possible to design more effective, problem-specific pruning rules that
substantially reduce the propagation cost from the very first layers; developing
such strategies for broad families of ans\"{a}tze remains an open problem.


\section{Applications}
\label{sec:applications}
\label{sec:apps}

The propagation method is applicable to any \gls{qml} problem whose loss can be
written as a function of expectation values of observables. We demonstrate the
approach on two benchmarks: \gls{vqe} on a two-dimensional transverse-field
Ising model (Sec.~\ref{subsec:vqe}), and quantum generative modeling of
binarized MNIST images~\cite{deng2012mnist} (Sec.~\ref{subsec:gen}). Both
experiments are conducted on a 64-qubit ($8\times8$) circuit, well beyond the
reach of exact statevector simulation, and the trained parameters are
subsequently deployed on the IBM quantum processor
\textit{ibm\_berlin}~\cite{ibmroadmap}.

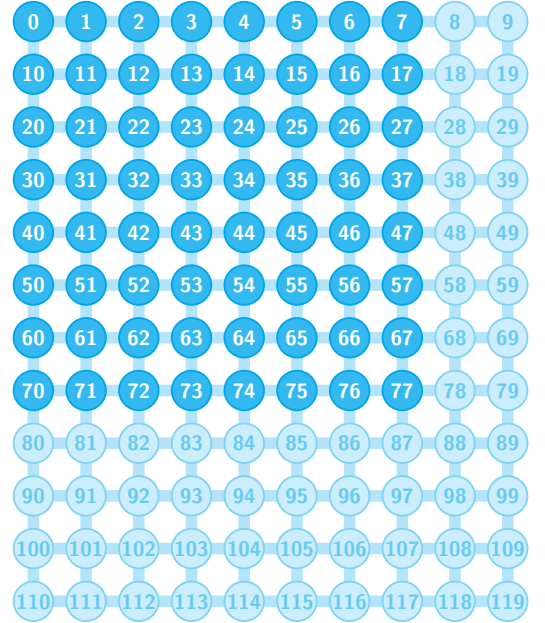
\begin{figure}[h]
	\centering
	\resizebox{.8\linewidth}{!}{
%
\begin{tikzpicture}[
		every node/.style={
				circle, draw, inner sep=0pt,
				minimum size=0.6cm,
				font=\sffamily\small\bfseries, text=secondarycolor!60,
			},
		lightnode/.style={fill=secondarycolor!20, draw=secondarycolor!50, line width=0.8pt},
		node/.style={fill=secondarycolor!80, draw=secondarycolor, line width=0.8pt, text=white},
		edge/.style={draw=secondarycolor!30, line width=5pt},
	]

	\def\dx{0.8}
	\def\dy{-0.8}

	\foreach \row in {0,...,11}{
			\foreach \col in {0,...,8}{
					\pgfmathtruncatemacro{\a}{\row*10 + \col}
					\pgfmathtruncatemacro{\b}{\row*10 + \col + 1}
					\draw[edge] (\col*\dx, \row*\dy) -- (\col*\dx+\dx, \row*\dy);
				}
		}
	\foreach \row in {0,...,10}{
			\foreach \col in {0,...,9}{
					\pgfmathtruncatemacro{\a}{\row*10 + \col}
					\pgfmathtruncatemacro{\b}{\row*10 + \col + 10}
					\draw[edge] (\col*\dx, \row*\dy) -- (\col*\dx, \row*\dy+\dy);
				}
		}

	\foreach \row in {0,...,11}{
			\foreach \col in {0,...,9}{
					\pgfmathtruncatemacro{\n}{\row*10 + \col}
					\pgfmathsetmacro{\x}{\col*\dx}
					\pgfmathsetmacro{\y}{\row*\dy}
					\ifnum\n=8 \node[lightnode] at (\x,\y) {\n};
					\else\ifnum\n=9 \node[lightnode] at (\x,\y) {\n};
						\else\ifnum\n=18 \node[lightnode] at (\x,\y) {\n};
							\else\ifnum\n=19 \node[lightnode] at (\x,\y) {\n};
								\else\ifnum\n=28 \node[lightnode] at (\x,\y) {\n};
									\else\ifnum\n=29 \node[lightnode] at (\x,\y) {\n};
										\else\ifnum\n=38 \node[lightnode] at (\x,\y) {\n};
											\else\ifnum\n=39 \node[lightnode] at (\x,\y) {\n};
												\else\ifnum\n=48 \node[lightnode] at (\x,\y) {\n};
													\else\ifnum\n=49 \node[lightnode] at (\x,\y) {\n};
														\else\ifnum\n=58 \node[lightnode] at (\x,\y) {\n};
															\else\ifnum\n=59 \node[lightnode] at (\x,\y) {\n};
																\else\ifnum\n=68 \node[lightnode] at (\x,\y) {\n};
																	\else\ifnum\n=69 \node[lightnode] at (\x,\y) {\n};
																		\else\ifnum\n=78 \node[lightnode] at (\x,\y) {\n};
																			\else\ifnum\n=79 \node[lightnode] at (\x,\y) {\n};
																				\else\ifnum\n=80 \node[lightnode] at (\x,\y) {\n};
																					\else\ifnum\n=81 \node[lightnode] at (\x,\y) {\n};
																						\else\ifnum\n=82 \node[lightnode] at (\x,\y) {\n};
																							\else\ifnum\n=83 \node[lightnode] at (\x,\y) {\n};
																								\else\ifnum\n=84 \node[lightnode] at (\x,\y) {\n};
																									\else\ifnum\n=85 \node[lightnode] at (\x,\y) {\n};
																										\else\ifnum\n=86 \node[lightnode] at (\x,\y) {\n};
																											\else\ifnum\n=87 \node[lightnode] at (\x,\y) {\n};
																												\else\ifnum\n=88 \node[lightnode] at (\x,\y) {\n};
																													\else\ifnum\n=89 \node[lightnode] at (\x,\y) {\n};
																														\else\ifnum\n=90 \node[lightnode] at (\x,\y) {\n};
																															\else\ifnum\n=91 \node[lightnode] at (\x,\y) {\n};
																																\else\ifnum\n=92 \node[lightnode] at (\x,\y) {\n};
																																	\else\ifnum\n=93 \node[lightnode] at (\x,\y) {\n};
																																		\else\ifnum\n=94 \node[lightnode] at (\x,\y) {\n};
																																			\else\ifnum\n=95 \node[lightnode] at (\x,\y) {\n};
																																				\else\ifnum\n=96 \node[lightnode] at (\x,\y) {\n};
																																					\else\ifnum\n=97 \node[lightnode] at (\x,\y) {\n};
																																						\else\ifnum\n=98 \node[lightnode] at (\x,\y) {\n};
																																							\else\ifnum\n=99 \node[lightnode] at (\x,\y) {\n};
																																								\else\ifnum\n=100 \node[lightnode] at (\x,\y) {\n};
																																									\else\ifnum\n=101 \node[lightnode] at (\x,\y) {\n};
																																										\else\ifnum\n=102 \node[lightnode] at (\x,\y) {\n};
																																											\else\ifnum\n=103 \node[lightnode] at (\x,\y) {\n};
																																												\else\ifnum\n=104 \node[lightnode] at (\x,\y) {\n};
																																													\else\ifnum\n=105 \node[lightnode] at (\x,\y) {\n};
																																														\else\ifnum\n=106 \node[lightnode] at (\x,\y) {\n};
																																															\else\ifnum\n=107 \node[lightnode] at (\x,\y) {\n};
																																																\else\ifnum\n=108 \node[lightnode] at (\x,\y) {\n};
																																																	\else\ifnum\n=109 \node[lightnode] at (\x,\y) {\n};
																																																		\else\ifnum\n=110 \node[lightnode] at (\x,\y) {\n};
																																																			\else\ifnum\n=111 \node[lightnode] at (\x,\y) {\n};
																																																				\else\ifnum\n=112 \node[lightnode] at (\x,\y) {\n};
																																																					\else\ifnum\n=113 \node[lightnode] at (\x,\y) {\n};
																																																						\else\ifnum\n=114 \node[lightnode] at (\x,\y) {\n};
																																																							\else\ifnum\n=115 \node[lightnode] at (\x,\y) {\n};
																																																								\else\ifnum\n=116 \node[lightnode] at (\x,\y) {\n};
																																																									\else\ifnum\n=117 \node[lightnode] at (\x,\y) {\n};
																																																										\else\ifnum\n=118 \node[lightnode] at (\x,\y) {\n};
																																																											\else\ifnum\n=119 \node[lightnode] at (\x,\y) {\n};
																																																												\else \node[node] at (\x,\y) {\n};
																																																												\fi\fi\fi\fi\fi\fi\fi\fi\fi\fi\fi\fi\fi\fi\fi\fi\fi\fi\fi\fi\fi\fi\fi\fi\fi\fi\fi\fi\fi\fi\fi\fi\fi\fi\fi\fi\fi\fi\fi\fi\fi\fi\fi\fi\fi\fi\fi\fi\fi\fi\fi\fi\fi\fi\fi\fi
				}
		}
\end{tikzpicture}} \caption{Qubit
		connectivity graph of the IBM quantum processor \textit{ibm\_berlin}
		used in this work ($12\times10$ grid, 120 qubits). Dark (blue) nodes are
		the 64 qubits (rows 0-7, columns 0-7) actively used in the experiment;
		light nodes are unused.}
	\label{fig:ibmberlin}
\end{figure}

\textit{ibm\_berlin} is a 120-qubit processor whose qubits are arranged in a
$12\times10$ square lattice, with each qubit coupled to up to four nearest
neighbors as shown in Fig.~\ref{fig:ibmberlin}. Our experiments use the
$8\times8$ sub-lattice spanning rows 0-7 and columns 0-7 of the full device.
Because gate errors and decoherence accumulate with circuit depth, minimizing
the number of two-qubit gates is critical for preserving result fidelity on
noisy intermediate-scale hardware.

\begin{figure}[h]
	\resizebox{\linewidth}{!}{
\begin{tikzpicture}[
		fashed/.style={
				thick,
				dash pattern= on 8pt off 0pt
				on 3pt off 1pt
				on 2pt off 1pt
				on 1pt off 1pt
			}
	]
	\node at (0, 0) {
		\begin{quantikz}[row sep={0.6cm,between origins}, wire types={q,q,q,n,q}]
			\lstick{$\ket{0}$} & \gate{RX} \gategroup[6,steps=7,style={dashed}, label style={label position=below,anchor=north,yshift=-0.2cm}]{} & \ctrl{1}    & \qw         & \gate{RX} & \control{} & \control{} & \gate{RX} & \gate{RY} & \\
			\lstick{$\ket{0}$} & \gate{RX}                                                                                                       & \control{1} & \ctrl{1}    & \gate{RX} & \control{} & \control{} & \gate{RX} & \gate{RY} & \\
			\lstick{$\ket{0}$} & \gate{RX}                                                                                                       & \ctrl{0}    & \control{1} & \gate{RX} & \control{} & \control{} & \gate{RX} & \gate{RY} & \\
			\lstick{\,\vdots}  & \vdots                                                                                                          & \vdots      & \vdots      & \vdots    &            &            & \vdots    & \vdots    & \\
			\lstick{$\ket{0}$} & \gate{RX}                                                                                                       & \ctrl{1}    & \ctrl{0}    & \gate{RX} & \control{} & \control{} & \gate{RX} & \gate{RY} & \\
			\lstick{$\ket{0}$} & \gate{RX}                                                                                                       & \control{1} & \qw         & \gate{RX} & \control{} & \control{} & \gate{RX} & \gate{RY} &
		\end{quantikz}};

	\draw[fashed](.6, 1.6) -- ++(-.4,-.4);
	\draw[fashed](.6, 1.0) -- ++(-.4,-.4);
	\draw[fashed](.6, .4) -- ++(-.4,-.4);
	\draw[fashed](.6, -.8) -- ++(-.4,-.4);
	\draw[fashed](.6, -1.4) -- ++(-.4,-.4);

	\draw[fashed](1.2, 1.6) -- ++(.4,.4);
	\draw[fashed](1.2, 1.0) -- ++(.4,.4);
	\draw[fashed](1.2, .4) -- ++(.4,.4);
	\draw[fashed](1.2, -.8) -- ++(.4,.4);
	\draw[fashed](1.2, -1.4) -- ++(.4,.4);
\end{tikzpicture}} \caption{Hardware-efficient
		ansatz: an $RX$ layer followed by a CZ entanglement layer is repeated
		twice: once for horizontal connections and once for vertical
		connections, and a final $RX$-$RY$ rotation layer. The gates inside the
		dashed block can be repeated multiple times, representing the
		\textit{depth} of the ansatz. The sketch shows a single row of the
		two-dimensional ansatz; the same pattern is repeated across all rows.}
	\label{fig:ansatz}
\end{figure}
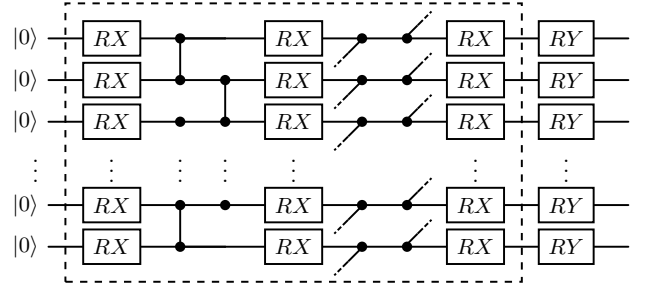

The hardware-efficient ansatz shown in Fig.~\ref{fig:ansatz} is designed with
this constraint in mind. Applied to the $8\times8$ grid, it consists of an
initial single-qubit $RX$ layer, a $CZ$ entangling layer over all horizontal
nearest-neighbor pairs, an intermediate $RX$ layer, a $CZ$ layer over all
vertical nearest-neighbor pairs, and a final $RX$-$RY$ rotation layer, for a
total of $256$ trainable parameters at \textit{depth} 1 and $384$ trainable
parameters at \textit{depth} 2 (corresponding of two repetitions of the gates in
the dashed block).

Both $CZ$ and $RX$ are native to \textit{ibm\_berlin}, minimizing transpilation
overhead and keeping the two-qubit gate depth at $4$ and $8$ for depth $1$ and
$2$ respectively. This hardware-efficient choice of the ansatz introduces a
rotation bias, so the circuit is not strictly \textit{locally scrambling}, the
condition under which the exponential suppression of high-weight contributions
is proven~\cite{classicallyestimating}. The bias is, however, only partial,
since, unlike a combination of $\mathit{CNOT}$ and $RX$ gates, the combination
of $CZ$ and $RX$ gates does not preserve the direction of Pauli components,
and the final $RY$ layer further mixes the directions. Moreover, the authors
of Ref.~\cite{classicallyestimating} showed numerically that low-weight Pauli
propagation remains accurate for practical, non-pathological circuits that
violate the locally scrambling assumption.

\subsection{Variational Quantum Eigensolver}
\label{subsec:vqe}

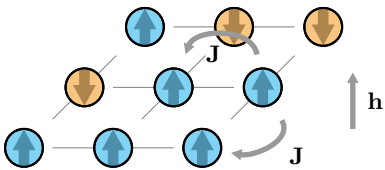
\begin{figure}[h]
	\centering
\newcommand{\spindown}{
	\begin{tikzpicture}[baseline=(c.center)]
		\node[draw=black, thick, circle, minimum size=.5cm, inner sep=0pt, fill=maincolor!50] (c) at (0,0) {};

		\draw[
		-{Triangle[width=8pt,length=8pt]},
		line width=4pt,
		color=maincolor!40!gray
		] (0, +.25) -- ++(0,-.5);

		\node[draw=black, thick, circle, minimum size=.5cm, inner sep=0pt] (c) at (0,0) {};

	\end{tikzpicture}
}
\newcommand{\spinup}{
	\begin{tikzpicture}[baseline=(c.center)]
		\node[draw=black, thick, circle, minimum size=.5cm, inner sep=0pt, fill=secondarycolor!50] (c) at (0,0) {};

		\draw[
		-{Triangle[width=8pt,length=8pt]},
		line width=4pt,
		color=secondarycolor!40!gray
		] (0, -.25) -- ++(0,+.5);

		\node[draw=black, thick, circle, minimum size=.5cm, inner sep=0pt] (c) at (0,0) {};

	\end{tikzpicture}
}
\begin{tikzpicture}[
	connection/.style={
	->,
	line width=2pt,
	black!40!white,
	opacity=1,
	>={Triangle[length=4, width=4]}
	},
	magfield/.style={
	->,
	line width=2pt,
	black!40!white,
	opacity=1,
	>={Triangle[length=4, width=4]}
	}
	]
	\pgfdeclarelayer{background}
	\pgfdeclarelayer{main}
	\pgfdeclarelayer{foreground}
	\pgfsetlayers{background,main,foreground}

	\begin{pgfonlayer}{main}
		\node (spin11) at (0, 0) {\spinup};
		\node (spin21) at ([xshift=+.8cm]spin11.east) {\spindown};
		\node (spin31) at ([xshift=+.8cm]spin21.east) {\spindown};

		\node (spin12) at ([xshift=-.8cm, yshift=-.8cm]spin11) {\spindown};
		\node (spin22) at ([xshift=+.8cm]spin12.east) {\spinup};
		\node (spin32) at ([xshift=+.8cm]spin22.east) {\spinup};

		\node (spin13) at ([xshift=-.8cm, yshift=-.8cm]spin12) {\spinup};
		\node (spin23) at ([xshift=+.8cm]spin13.east) {\spinup};
		\node (spin33) at ([xshift=+.8cm]spin23.east) {\spinup};

		\draw[magfield] (2.75, -1.35) -- coordinate (mag) ++(0, 0.75);
		\node at ([xshift=.3cm]mag) {$\mathbf{h}$};

		\draw[connection]
		([yshift=+.2cm]spin32) to[out=100, in=70] (spin22);
		\draw[connection]
		([yshift=+.2cm]spin32) to[out=300, in=-10] (spin33);

		\node at ([xshift=-.65cm, yshift=.45cm]spin32) {$\mathbf{J}$};
		\node at ([xshift=.45cm, yshift=-.9cm]spin32) {$\mathbf{J}$};

	\end{pgfonlayer}

	\begin{pgfonlayer}{background}
		\draw[gray, line width=0.1pt] (spin11) -- (spin31);
		\draw[gray, line width=0.1pt] (spin12) -- (spin32);
		\draw[gray, line width=0.1pt] (spin13) -- (spin33);

		\draw[gray, line width=0.1pt] (spin11) -- (spin13);
		\draw[gray, line width=0.1pt] (spin21) -- (spin23);
		\draw[gray, line width=0.1pt] (spin31) -- (spin33);

	\end{pgfonlayer}
\end{tikzpicture}
	\caption{Illustration of the 2D Ising spin model on a $3\times3$ lattice.
		Filled circles represent spins, with nearest-neighbor coupling
		$\mathbf{J}$ and external transverse field $\mathbf{h}$.}
	\label{fig:ising}
\end{figure}

\gls{vqe}~\cite{vqe,dualrolelowweightpauli} estimates the ground-state energy of
a target Hamiltonian by minimizing the variational energy
$\langle\psi(\boldsymbol\theta)|H|\psi(\boldsymbol\theta)\rangle$ over the
circuit parameters $\boldsymbol\theta$. Other works have instead warm-started
\gls{vqe} by initializing the ansatz with a classically pre-computed approximate
complex amplitude encoding of the ground state~\cite{truger2024warm}; the
resulting circuit must still be executed and refined on quantum hardware, and
the cost of constructing the encoding itself grows with the system size. Pauli
propagation instead removes quantum hardware from the training loop altogether:
the cost of optimization is governed solely by the weight and trigonometric-degree cutoffs
$w_\mathrm{cut}, \delta_\mathrm{cut}$ rather than by the number of qubits, allowing
it to scale to the 64-qubit setting considered here. We consider the
two-dimensional transverse-field Ising model (Fig.~\ref{fig:ising}) on the
$8\times8$ qubit lattice, with nearest-neighbor coupling $J$ and transverse
field $h$:

\begin{equation}
	\label{eq:isingH}
	H(J, h) = -\frac{1}{N}\Big(J\sum_{\langle i,j \rangle} Z_i Z_j + h\sum_i X_i\Big)
\end{equation}

Because $H$ decomposes into two independent Pauli sums

\begin{equation}
	\label{eq:isingHprop}
	H(J, h) = -\frac{J}{N}\Big(\underbrace{\sum_{\langle i,j \rangle} Z_i Z_j}_{O_1}\Big) -\frac{h}{N}\Big(\underbrace{\sum_i X_i}_{O_2}\Big)
\end{equation}

the \gls{vqe} loss requires propagating only two observables through the ansatz

\begin{align}
	\mathcal{L}_\text{VQE}(\vec{\theta}) \;=\; & \langle \psi(\vec\theta) | H | \psi(\vec\theta) \rangle \notag                               \\
	\;=\;                                      & -\frac{J}{N} \langle O'_1(\vec\theta) \rangle - \frac{h}{N} \langle O'_2(\vec\theta) \rangle
	\label{eq:isingloss}
\end{align}

and by linearity, only one propagation is required for any possible value of $J$
and $h$.

\begin{algorithm}
	\caption{VQE via Pauli Propagation}
	\label{alg:vqe}
	\KwIn{Ansatz $U(\boldsymbol{\theta})$, Hamiltonian $H$, cutoffs
		$w_\mathrm{cut}, \delta_\mathrm{cut}$, learning rate $\eta$, steps $T$}
	\KwOut{Optimized parameters $\boldsymbol{\theta}^*$}

	Build Propagator $\mathcal{P}$ for
	$f(\boldsymbol{\theta})=\langle\psi(\boldsymbol{\theta})|H|\psi(\boldsymbol{\theta})\rangle$
	and propagate with cutoffs $(w_\mathrm{cut}, \delta_\mathrm{cut})$\;

	Initialize $\boldsymbol{\theta}_0 \sim \mathcal{U}(-\pi, \pi)$\;

	\For{$t = 1$ \KwTo $T$}{ Evaluate $f(\boldsymbol{\theta}_{t-1})$ and
		$\nabla_{\boldsymbol{\theta}} f$ analytically\; $\boldsymbol{\theta}_t
			\leftarrow
			\mathrm{Adam}\!\left(\boldsymbol{\theta}_{t-1},\,\nabla_{\boldsymbol{\theta}}
			f,\,\eta\right)$\; } \Return{$\boldsymbol{\theta}_T$}
\end{algorithm}

\begin{figure}[h]
	\centering
	\includegraphics{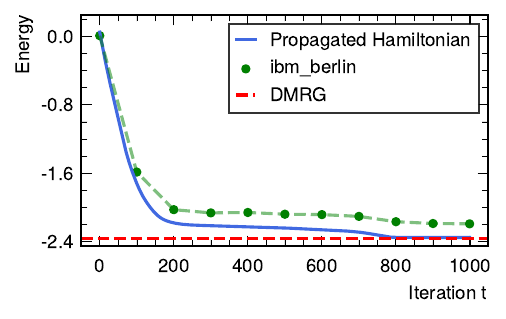}
	\caption{VQE energy as a function of training iteration for the depth-1
		ansatz on the 2D Ising Hamiltonian at $(J, h) = (1, 2)$. The red dashed
		line marks the DMRG ground-state energy, serving as the theoretical
		lower bound. The solid blue line shows the training loss under exact
		Pauli propagation, with no weight or trigonometric-degree cutoff. The green
		dashed line and markers show the energy evaluated on
		\textit{ibm\_berlin} at the corresponding training iterations.}
	\label{fig:vqeresult}
\end{figure}

\begin{figure}[h]
	\centering
	\includegraphics{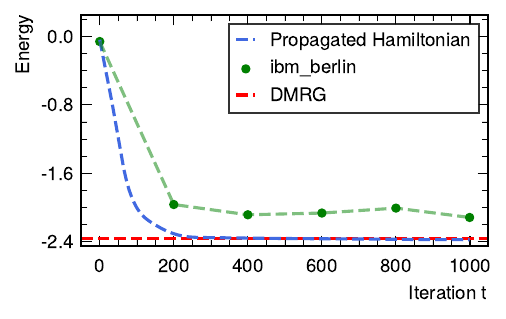}
	\caption{VQE energy as a function of training iteration for the depth-2
		ansatz on the 2D Ising Hamiltonian at $(J, h) = (1, 2)$. The red dashed
		line marks the DMRG ground-state energy. The dashed blue line shows
		the training loss under Pauli propagation with a weight cutoff
		$w_\mathrm{cut} = 12$ and no trigonometric-degree cutoff, that is, an
		approximated expectation-value function. The green dashed line and
		markers show the energy evaluated on \textit{ibm\_berlin} at the
		corresponding training iterations.}
	\label{fig:vqeresult2}
\end{figure}

Figures~\ref{fig:vqeresult} and~\ref{fig:vqeresult2} report training and
hardware results at $J = 1$, $h = 2$, placing the system in the
ferromagnetic phase ($h < h_c \approx 3.044\,J$), close to the quantum
critical point. We train two ans\"atze of depth 1 and depth 2 respectively,
both for $T = 1000$ steps with the Adam optimizer~\cite{adam} at learning
rate $\eta = 0.005$, starting from parameters drawn uniformly in $[-\pi, \pi]$.
The full training procedure is detailed in Algorithm~1. The reference
ground-state energy is obtained with DMRG on a snake-path matrix product state
with maximal bond dimension $\chi = 256$ and $15$ sweeps.
For the depth-1 ansatz, the Hamiltonian is propagated exactly, without
weight or trigonometric degree truncation. For the depth-2 ansatz, exact propagation is
no longer tractable and a weight cutoff $w_\mathrm{cut} = 12$ is required,
with no trigonometric-degree cutoff. The resulting training loss is therefore an
approximation of the true expectation-value function, shown as a dashed
line in Figure~\ref{fig:vqeresult2}, as opposed to the solid, exact curve in
Figure~\ref{fig:vqeresult}. At several iterations during training we also
evaluate the energy on \textit{ibm\_berlin}, transpiling the circuits with
the Qiskit preset pass manager at optimization level 1 and using the Qiskit
Runtime Estimator at resilience level 2, which combines readout-error mitigation
with zero-noise extrapolation~\cite{temme2017error,kandala2019error}, and
$4096$ shots. In both cases a persistent gap separates the hardware curve
from the propagated-Hamiltonian curve throughout training. Since this gap is
already present for the depth-1 ansatz, where propagation is exact, we
attribute it to the hardware's precision in estimating the expectation
value of this observable rather than to the training procedure itself.
Training at depth 2 converges in fewer iterations than at depth 1, but the
deployed energy on \textit{ibm\_berlin} is slightly worse.

One important practical benefit of training classically is that quantum
hardware is not required during optimization. The entire iterative
procedure, all $1000$ Adam steps with their gradient evaluations, runs on a
classical computer for both ansatz depths. To track how closely the
deployed model follows the classically propagated energy over the course of
training, we additionally query \textit{ibm\_berlin} at several checkpoint
iterations for each model, evaluating $\langle H \rangle$ with $4096$ shots
at resilience level 2 each time. These hardware evaluations are not part of
the optimization loop and serve only to produce
Figures~\ref{fig:vqeresult} and~\ref{fig:vqeresult2}. In a conventional
\gls{vqe} workflow every training iteration would require a circuit
execution, so bringing training fully to the classical side reduces quantum
usage to only what is needed for validation.

\subsection{Generation}
\label{subsec:gen}

Quantum generative models of handwritten digits have previously been trained
and sampled on trapped-ion hardware~\cite{rudolph2022generation}, and quantum
generative models are being actively explored for scientific applications such
as calorimeter shower simulation in high-energy
physics~\cite{ciqp,qfan,qpce,comb}.
Mirroring the work in~\cite{trainonclassical}, it is possible to train a
sampling-type quantum generative model through a loss function of Pauli-Z
expectation values:

\begin{equation}
	\label{eq:genloss}
	\mathcal{L}_\text{MMD}(\vec{\theta}) \;=\; \frac{1}{|\mathcal{A}|}\sum_j\left(
	\langle Z_{a_j}\rangle_\text{data} -
	\langle Z_{a_j}(\vec{\theta})\rangle
	\right)^2
\end{equation}

where $\mathcal{A} = \{a_j \sim \mathcal{P}_\sigma\}$. Given the large number
of observables that must be propagated at every round, the ansatz depth is
kept at $1$, so that each propagation remains fast enough to be repeated
many times during training.

This formula was originally derived in~\cite{trainabilitymmd}, and it represents
an estimator of the squared \gls{mmd}~\cite{gretton2012kernel} (MMD$^2$) with Gaussian kernel
at bandwidth $\sigma$. Namely, it consists in matching the $Z$ expectation
values of the trained circuit sampled from the Bernoulli distribution
parametrized by $\sigma$ to the equivalent expectation values from the training
data, when this loss is minimized, the output state will be in a superposition
of all its possible samples, obtained without ever accessing the full state
vector.

\begin{algorithm}
	\caption{MMD Training}
	\KwIn{Ansatz $U(\boldsymbol{\theta})$, training set $\mathcal{X}$, bandwidth
		$\sigma$, cutoffs $w_\mathrm{cut}, \delta_\mathrm{cut}$, learning rate
		$\eta$, rounds $R$, steps per round $S$} \KwOut{Optimized parameters
		$\boldsymbol{\theta}^*$}

	Initialize $\boldsymbol{\theta} \sim \mathcal{N}(0, 1)$\;

	\For{$r = 1$ \KwTo $R$}{ Sample a batch $\{O_j\}$ of Pauli-$Z$ observables
		from a binomial distribution parametrized by $\sigma$\;

		Compute the dataset's expectation values $\{\langle
			O_j\rangle_\text{data}\}$

		Propagate $\{O_j\}$ analytically through $U(\boldsymbol{\theta})$ in the
		Heisenberg picture with cutoffs $(w_\mathrm{cut}, \delta_\mathrm{cut})$\;

		\For{$s = 1$ \KwTo $S$}{ Evaluate
			$\mathcal{L}_\text{MMD}(\boldsymbol{\theta}) =
				\frac{1}{|\mathcal{A}|}\sum_k \bigl(\langle
				O_i\rangle_{\boldsymbol{\theta}} - \langle
				O_i\rangle_\text{data}\bigr)^2$ and $\nabla_{\boldsymbol{\theta}}
				\mathcal{L}_\text{MMD}$ analytically\; $\boldsymbol{\theta}
				\leftarrow
				\mathrm{Adam}(\boldsymbol{\theta},\,\nabla_{\boldsymbol{\theta}}
				\mathcal{L}_\text{MMD},\,\eta)$\; } }
	\Return{$\boldsymbol{\theta}$}
\end{algorithm}

\begin{figure}
	\centering

	\resizebox{\linewidth}{!}{
		\includegraphics{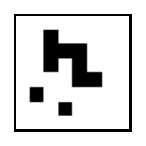}
		\includegraphics{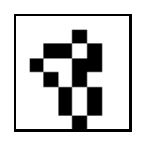}
		\includegraphics{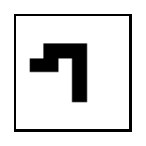}
		\includegraphics{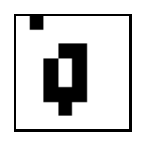}
		\includegraphics{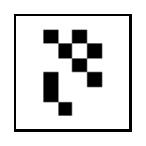}
	}

	\resizebox{\linewidth}{!}{
		\includegraphics{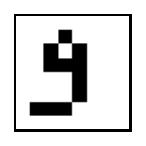}
		\includegraphics{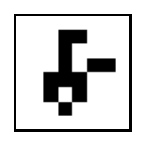}
		\includegraphics{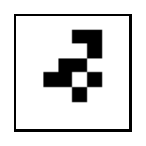}
		\includegraphics{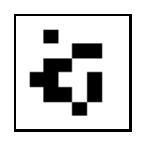}
		\includegraphics{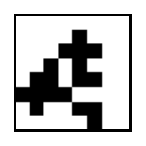}
	} \caption{Samples generated by measuring the trained circuit on the IBM
		quantum processor \textit{ibm\_berlin}. Each pixel corresponds to a
		qubit measurement outcome, represented as white (state $\ket{0}$) or
		black (state $\ket{1}$).}
	\label{fig:gen_samples}
\end{figure}

\begin{table}
	\begin{ruledtabular}
		\begin{tabular}{lcc}
			Model                  & Parameters & Test MMD$^2$ ($\times10^{-3}$) \\
			\hline
			Propagator             & $256$      & $\mathbf{2.05 \pm 0.18}$       \\
			IQP ($1\times$)        & $256$      & $24.11 \pm 1.16$               \\
			IQP ($10\times$)       & $2560$     & $2.51 \pm 0.22$                \\
			\hline
			Train-vs-test baseline & --         & $0.31 \pm 0.01$                \\
		\end{tabular}
	\end{ruledtabular}
	\caption{Test MMD$^2$ of the generative models, reported as mean $\pm$
		standard deviation over $100$ independent draws of $1000$ test
		observables. The train-vs-test baseline is the score of a model
		reproducing the training moments exactly. Lower MMD$^2$ indicates a
		better match to the target distribution.}
	\label{tab:gen_compare_iqp}
\end{table}

\begin{figure}
	\centering
	\includegraphics{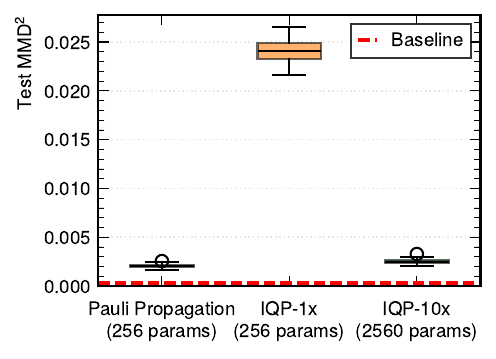}
	\caption{Distribution of the test MMD$^2$ over the $100$ observable draws
		for the propagator and the two IQP baselines summarized in
		Table~\ref{tab:gen_compare_iqp}. The dashed red line marks the
		train-vs-test baseline.}
	\label{fig:gen_compare_iqp}
\end{figure}

Unlike the \gls{vqe} setting, where propagating two observables is easily
tractable, the \gls{mmd} estimator requires propagating a fresh batch of
$|\mathcal{A}| = 1000$ observables at every round, making exact propagation
prohibitively expensive. The training data consist of $5000$ MNIST images of
all ten digits, downsampled to $8\times8$ pixels with Lanczos interpolation and
binarized with threshold $0.3$. The bandwidth $\sigma$ is set via the median
heuristic on $100$ training images, giving $\sigma \approx 3.61$, and weight
truncation is applied to keep the per-round cost manageable. The circuit is
trained for $R = 100$ rounds of $S = 1000$ Adam steps each, at learning rate
$\eta = 5\times10^{-5}$, with weight cutoff $w_\mathrm{cut} = 12$ and no
trigonometric-degree cutoff, for a total of $100{,}000$ optimization steps,
starting from parameters drawn from $\mathcal{N}(0, 1)$; the full
procedure is detailed in Algorithm~2. Figure~\ref{fig:gen_samples}
shows samples obtained by deploying the trained ansatz on \textit{ibm\_berlin}
(transpiled at optimization level 1) and performing single-shot
measurements of all 64 qubits, with each shot yielding one image. Since each
sample is a single measurement outcome, no error mitigation is applied. A handful of samples clearly reproduce recognizable digit
shapes, while others display a central indistinguishable mesh. Several factors
contribute to this: the shallow ansatz may lack the expressivity to fully
capture the target distribution; the MMD loss itself is trained only
approximately, as the high cost of propagating many observables per round
necessitates truncation and limits the number of training rounds; and finally,
the single-shot nature of the output makes the images highly sensitive to
hardware noise as a single bit-flip on any qubit corrupts the corresponding
pixel and can render an otherwise correct sample unrecognizable.


Just as in the \gls{vqe} case, the quantum device is used only at the very end
of the pipeline. For all the training the gradient evaluations and the Adam
updates, run on classical hardware. Once the parameters are fixed, a single
circuit execution of 64 shots on \textit{ibm\_berlin} yields 64 generated images
in roughly 2 seconds. The quantum usage time is thus reduced to the minimum.

\begin{figure*}
	\centering
	\resizebox{\linewidth}{!}{\includegraphics{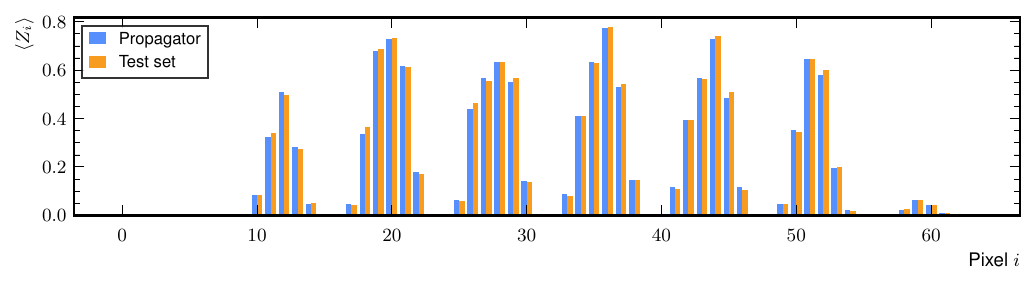}}\\[-.3cm]
	\caption{First moment of the output distribution: per-qubit average $\langle
			Z_i \rangle$ evaluated on the trained circuit (blue) versus the test set
		(orange). The model reproduces the marginal single-site statistics of
		the data with high fidelity.}
	\label{fig:gen_firstmom}
\end{figure*}

\begin{figure*}
	\centering
	\includegraphics{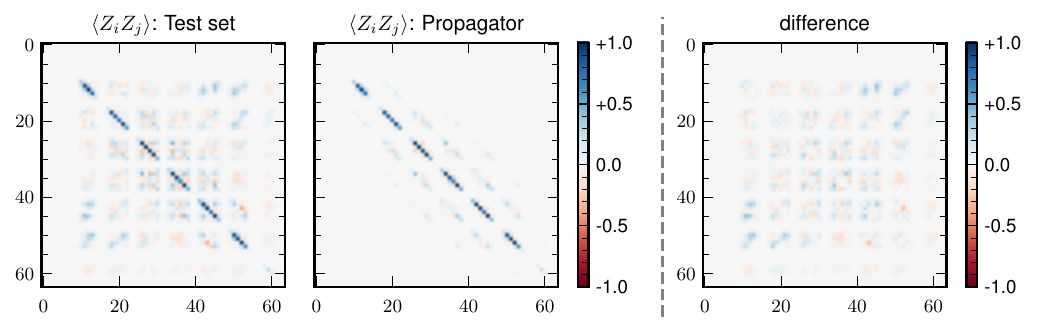}
	\caption{Second moment of the output distribution: connected two-site
		correlators $\langle Z_i Z_j\rangle - \langle Z_i\rangle\langle
			Z_j\rangle$ evaluated on the test set (left) and on the trained circuit
		(middle). Agreement is less tight than for the first moment, indicating
		that capturing pairwise correlations exactly remains a harder task for
		the shallow ansatz.}
	\label{fig:gen_secondmom}
\end{figure*}

Beyond the visual inspection of individual samples, we can quantify how well the
trained distribution matches the data by comparing its statistical moments
against those of the test set. Figures~\ref{fig:gen_firstmom}
and~\ref{fig:gen_secondmom} show this comparison for the first and second
central moments. Rather than drawing samples, both quantities are computed
directly from quantum observables: the first moment from the single-site
expectation values $\langle Z_i \rangle$, and the second central moment from the
connected two-point correlators $\langle Z_i Z_j\rangle - \langle
	Z_i\rangle\langle Z_j\rangle$. The first-moment results
(Fig.~\ref{fig:gen_firstmom}) are quite promising: the per-qubit marginals of
the model closely track those of the test set, indicating that the single-site
statistics of the learned distribution are well captured. The second-moment (Fig.~\ref{fig:gen_secondmom})
however does not capture the off-diagonal terms, meaning that the model struggled
to capture the higher-order statistics of the data.

Table~\ref{tab:gen_compare_iqp} and Fig.~\ref{fig:gen_compare_iqp} compare the test MMD$^2$ (computed by evaluating
the squared MMD on a set of observables not present in the training set) of the trained
propagator against two IQP circuit baselines~\cite{iqpopt} with equal and
$10\times$ more parameters, respectively. Both IQP circuits are built from
gates drawn from a nearest-neighbor lattice generator set on the $8\times8$
grid, with interaction distance $2$ and maximum Pauli weight $4$, keeping
$256$ gates chosen at random for the equal-parameter circuit and $2560$ for
the $10\times$ circuit. Both IQP circuits are initialized from the
data~\cite{iqpopt} and trained on the same $5000$ training images with the
same optimization budget as the propagator, namely $100{,}000$ Adam steps at
learning rate $\eta = 5\times10^{-5}$, estimating the MMD$^2$ loss with $1000$
observables and $1000$ samples per observable at each step, using the same
median-heuristic bandwidth. The test MMD$^2$ is evaluated on the full MNIST test set
($10{,}000$ images) over $100$ independent draws of $1000$ observables each,
shared by all models. For the IQP circuits the expectation values are
estimated with $10{,}000$ samples per observable, while the expectation values of the trained
ansatz are computed with the truncated propagator ($w_\mathrm{cut} = 12$)
rather than exactly. On the held-out test set, the propagator reaches
a mean test MMD$^2$ of $0.00205 \pm 0.00018$, against $0.02411 \pm 0.00116$ for
the equal-parameter IQP circuit and $0.00251 \pm 0.00022$ for the $10\times$
circuit. As a reference, the last row of Table~\ref{tab:gen_compare_iqp}
(dashed line in Fig.~\ref{fig:gen_compare_iqp}) reports the train-vs-test baseline, obtained by replacing the model expectation
values with the empirical moments of the training images and evaluating the
same MMD$^2$ estimator, with the same held-out images and observable draws,
against the test set. Since training targets exactly these moments, this is
the score of a model that fits its training data perfectly, and it yields
$0.00031 \pm 0.00001$, well below all three models (the finite-sample noise
floor of the test set alone is $\approx 0.00005$), so the differences
between the models reflect genuine modelling error rather than evaluation
noise. The propagator achieves a roughly $12$ times lower MMD$^2$ than the
equal-parameter IQP baseline, and also outperforms the $10\times$ baseline
despite using a tenth of the parameters, scoring lower on every one of the
$100$ observable draws. This parameter
efficiency is not entirely surprising: in IQP circuits all gates commute, so
the circuit applies a set of independent transformations with no interaction
among parameters (analogous to a linear layer in a classical neural network).
In contrast, the non-commuting gates of our ansatz introduce a causal
structure in which each gate's effect propagates through subsequent layers,
making every parameter collectively more expressive. This structural
advantage also translates directly to shallower circuits that remain
feasible to deploy on real quantum hardware.

\section{Discussions and conclusions}
\label{sec:conc}

We have presented Pauli propagation as a practical method for classically
pre-training large shallow \glspl{pqc} on \gls{qml} tasks whose loss functions
are expressible as observable expectation values. The results on both the 2D
Ising \gls{vqe} and the MNIST generative model are promising and, importantly,
constitute a proof of concept rather than a claim of optimality: the primary
goal was to demonstrate that the train-on-classical, deploy-on-quantum pipeline
is viable at 64 qubits, and both experiments confirm this.

Whether Pauli propagation can establish itself as a general-purpose alternative
training method remains an open question that warrants further investigation. A
key motivation for pursuing this direction is that fully offloading training to
classical resources may ultimately be the only scalable path forward: as qubit
counts grow into the hundreds and beyond, gradient estimation on hardware
becomes prohibitively expensive, and classical surrogates such as the one
proposed here are among the few approaches that scale without demanding repeated
quantum circuit evaluations.

One promising avenue for improving hardware results is to incorporate device
noise directly into the propagation. Pauli propagation has recently been
extended to handle local noise channels~\cite{angrisani2025simulating},
including non-unital noise~\cite{martinez2025nonunital}, which
would allow the classical optimizer to account for the noise model of the target
device and train parameters that are robust to, rather than degraded by,
realistic error rates.

One limitation surfaced by the generative experiment is that the MMD loss, while
directly expressible as observable expectation values and therefore compatible
with Pauli propagation, is not the most efficient objective for this workflow.
Matching the expectation values of many Pauli-Z strings requires propagating a
large number of observables per training round, a cost that grows with the
desired statistical accuracy and stands in contrast to IQP-based approaches,
where the evaluation of expectation values of many different observables can be
performed at a low cost. Identifying loss functions that are both physically
motivated and cheaper to propagate, remains an open direction for future work on
classically trained quantum generative models. Other training objectives have
been proposed for classically trainable quantum generative models, such as the
log-likelihood loss based on Parseval's identity introduced
in~\cite{quantumfourier}. It remains worth investigating whether training
objectives other than the \gls{mmd} estimator of
Eq.~\eqref{eq:genloss} could benefit better from the Pauli propagation framework.

A further opportunity opened by this framework is the ability to inspect an
explicit (exact or approximated) functional form of the loss before running a
single quantum experiment. Such access could yield insights that are otherwise
difficult to obtain. Studying how the propagated expression degenerates with
depth and connectivity could clarify the onset of barren plateaus~\cite{areviewofbarren}, pointing
toward a loss-informed way of constructing ans\"{a}tze that avoid them. Along
the same lines, the explicit dependence on the parameters could explain why
small-angle initializations are empirically effective, and guide the search for
alternative initialization schemes that yield higher gradient variance from the
start. Finally, the same expression directly exposes the relative importance of
each parameter to the loss, which could be used to identify and prune parameters
that contribute negligibly, reducing circuit depth without sacrificing
performance. No systematic methodology in synergy with this method exists yet,
and developing one represents a compelling direction for future work.

\section{Acknowledgments}
This research was supported in part through the Maxwell computational resources
operated at Deutsches Elektronen-Synchrotron DESY (Hamburg, Germany), a member
of the Helmholtz Association HGF. ENGAGE has received funding from the European
Union’s Horizon 2020 Research and Innovation Programme under the Marie
Skłodowska-Curie Grant Agreement No. 101034267. We thank Florian Rehm for the useful
conversations and suggestions and Francesco di Marcantonio for the DMRG analysis.\\

\section{Code availability}
All code developed and used for the analyses presented in this work is publicly
available on GitHub at
\url{https://github.com/SaverioMonaco/Pauli-Propagator}~\cite{githubrepo}.

\nocite{kay2018tutorial, pennylane, qiskit2024}
\bibliography{references}

\end{document}